\documentclass[review]{elsarticle}

\usepackage{hyperref}
\usepackage{amsmath}
\journal{Journal of \LaTeX\ Templates}

\begin{document}
\begin{frontmatter}

\title{Dynamics of market states and risk assessment}
\author[1,4]{Hirdesh K. Pharasi\corref{mycorrespondingauthor}}
\author[2]{Eduard Seligman}
\author[3,4]{Suchetana Sadhukhan}
\cortext[mycorrespondingauthor]{Corresponding author}
\author[4,5]{Parisa Majari}
\author[4,5]{Thomas H. Seligman}


\address[1]{School of Engineering and Technology, BML Munjal University, Gurugram, Haryana-122413, India}
\address[2]{Swiss Air Trainer S.A. Payerne, Vaud, Switzerland}
\address[3]{Department of Physics, School of Advanced Sciences and Languages, VIT Bhopal, Kothri Kalan, Sehore-466114, India}
\address[4]{Instituto de Ciencias F\'{i}sicas, Universidad Nacional Aut\'{o}noma de M\'{e}xico, Cuernavaca-62210, M\'{e}xico}
\address[5]{Centro Internacional de Ciencias, Cuernavaca-62210, M\'{e}xico}

\begin{abstract}
Previous research explored various conditions of financial markets based on the similarity of correlation structures and classified as market states. We introduce modifications to previous selection criteria for these market states, mainly due to increased attention to the transition matrix between the states. Clustering and thus market states are fixed by the optimization of two parameters-- number of clusters and noise suppression, but in similar conditions, we give preference to the clustering which avoids large jumps in the transition matrix. We found statistically significant results applying this model to the S\&P 500 and Nikkei 225 markets for the pre-COVID-19 pandemic era (2006-2019). Retaining the epoch length of 20 trading days but reducing the shift of the epoch to a single trading day we are led to the concept of a trajectory of the market in the space of correlation matrices. We may visualize these states after dimensional scaling to two or three dimensions. This approach, using dynamics, improves the options of risk assessment, opens the door to dynamical treatments of markets (e.g. hedging), and shows noise suppression in a new light.
\end{abstract}


\end{frontmatter}

\section{Introduction}

Stock markets are examples of evolving complex systems~\cite{Bouchaud_2003,Mantegna_2007}. The efforts to understand the dynamics and statistical properties of a market based on the correlation between stocks are attracting researchers from different fields~\cite{Mantegna_2007,laloux_2000,goetzmann2001long,alexander2001market,tumminello2010correlation,Chakraborti_2011a,Chakraborti_2011b, junior2012correlation,chiang2007dynamic, chakraborti2020emerging, youssef2021dynamic, so2021impacts,zhang2020financial,sarwar2023market}. The health of a financial market can be examined by the amount of correlation between its constituent stocks. During market crashes the correlation is higher than during \textit{business as usual} periods. 

Some years ago the concept of market states (MS) based on similarity of the correlation matrices was put forward for the first time by M\"unnix et al.~\cite{Munnix_2012}, though the idea of market states was not entirely new to the financial literature~\cite{schaller1997regime, marsili2002dissecting}. It has a fairly wide acceptance, including some recent works by our group~\cite{Pharasi_2018, Pharasi_2019,chetalova2015zooming,stepanov2015stability,
rinn2015dynamics, guhr2020exact, Heckens_2020, Meudt_2015,guhr2015non,chetalova2015dependence} where we proposed alternate clustering criteria, which resulted in a smaller and probably less arbitrary number of market states. Yet interaction with the financial industry showed that while they were interested in the basic idea, the inclusion of the last century data seems inappropriate and also epochs shifted by $10$ trading days did not look attractive, as trading and hedging decisions have to be taken fast. We therefore have shifted the movement of the epoch to a single day, which is the shortest time span one can use without getting into the almost separate field of intraday trading and the necessity of obtaining intraday data. 

We then use multidimensional scaling technique on the distance matrix calculated as a measure of similarities among the correlation matrices constructed from one day shifts and apply the $k$-means clustering. For robust clustering, we follow the work by \cite{Pharasi_2018} and simultaneously choose an optimal cluster-number and  power for the noise-reducing power map \cite{Vinayak_2013}.  Our judgment for the cutoff is based on the coincidence of several runs of the $k$-means method, initiated from different initial random conditions, using the average cluster radius as criterion. The power map, though little used, is an efficient method to suppress noise directly in the correlation matrix~\cite{Guhr_2003, Schafer_2010}, but also to lift the degeneracy of the zero modes associated with short time series~\cite{Vinayak_2013,chakraborti2020emerging,damgaard2010microscopic}. Following Ref.~\cite{Pharasi_2018}, we calculate the transition matrix and find that using daily data makes this matrix nearly tridiagonal for the S\&P 500 market thus confirming the cluster choice made; we can go one step further and add tridiagonality to the above selection criteria of choice for the number of clusters and noise reduction parameter if the power used in power map is not considered. The practical usefulness of the method is, we believe, considerably enhanced by these changes.

In the next section, we describe in detail the stocks from the indices, the S\&P 500 and the Nikkei 225, for the time-span 2006-2019. We also specify the different steps taken to display and analyze the data in section~\ref{method}. In section~\ref{results} we show and discuss the clustering results first for the S\&P 500 and then for the Nikkei 225. In the first we proposed five states arranged linearly (roughly along x-axis) according to the average of the elements of the correlation matrix  representing each state. In the later, for the Nikkei 225 we opt for a seven state representation. We find that the stochastic dynamics, as represented by the transition matrix between the states, has a bifurcation between states 2 and 3, and  4 and 5 displaying similar average correlations. Also transitions between these two pairs of bifurcated states are few. This is a clear sign that the Japanese market obeys much more complicated dynamics.  Yet the central fact for risk assessment remains as transitions occur to the critical state 5 from state 4 for S\&P 500 market and to state 7 exclusively from state 6 for Nikkei 225 market. It is followed by a  discussion of the results and their risk implications with a general conclusion and outlook in section~\ref{discussions}.
                                                                                 
\section{Data description}\label{datadis}
We used the adjusted daily closure prices of the stocks from two indices, namely the S\&P 500 for US market (USA) and Nikkei 225 for the Japanese market (JPN)~\cite{Yahoo_finance}. Next we considered $N = 350$ stocks from the S\&P 500 index and $N = 156$ stocks of the Nikkei 225 index traded in the 14-year period from January 2006 to December 2019 which correspond to $T=3523$ and $T = 3459$ trading days, respectively. Here we included only those stocks which were present for the entire period of 14 years. Out of these stocks, we also filtered out a few stocks with missing data points of more than two consecutive trading days. The stock value for the missing days was chosen to be the same as the one for the last day with a closing entry. The stocks considered for the analysis are listed in Table SI and SII of the supplementary material. 
\section{Methods}\label{method}  
We start from the basic idea of market state of a financial market based on the similarity of correlation matrices at different times \cite{Munnix_2012,Pharasi_2018,Pharasi_2019,chetalova2015zooming,stepanov2015stability,
rinn2015dynamics, guhr2020exact,Heckens_2020,Meudt_2015,guhr2015non}. We use the time series of the daily adjusted closing prices to obtain a data matrix for stocks. With these data we follow the course of Ref.~\cite{Pharasi_2018} and use logarithmic increments $r_i(\tau)$ of each stock $i$ to obtain the relevant time series, where $\tau$ is the end date of the corresponding epoch. This allows us to construct the elements of the correlation matrix $\rho(\tau)$ for any given epoch within the time horizon as a Pearson correlation \cite{sharma2005text}: 
\begin{equation}
\rho_{ij}(\tau) = \frac{{\langle r_i r_j \rangle - \langle r_i \rangle \langle r_j \rangle}}{{\sqrt{\langle r_i^2\rangle - \langle r_i\rangle^2} \sqrt{\langle r_j^2 \rangle- \langle r_j \rangle^2}}}
\end{equation}
where the epoch averages $\langle \dots \rangle $ are computed over epochs of size $M=20$ trading days with $i,j=1,...,N$. Note that we do not compensate for weekends or holidays! In case of short time series the correlation matrices become highly singular~\cite{laloux_1999,Plerou_1999}. We use the power map method~\cite{Pharasi_2018,Pharasi_2019,Guhr_2003,vinayak_2014}, to reduce the noise of the short time correlation matrices. In this method, a nonlinear distortion is given to each cross-correlation coefficient within an epoch by: $C_{ij} = sign(\rho_{ij})|\rho_{ij}|^{1+\epsilon}$ , where $\epsilon \in (0,1)$ is the noise-suppression parameter. We thus study the evolution of the noise reduced cross-correlation matrices $C(\tau)$ of logarithmic returns of stocks for epochs of $M=20$ days and shifts of $\Delta=1$ day throughout the time horizon $2006-2019$. 

We now define similarity measure or distance between two correlation matrices $C(\tau_1)$ and $C(\tau_2)$ evaluated at different time $\tau_1$ and $\tau_2$ by the distance: 
\begin{equation}
\zeta(\tau_1,\tau_2) \equiv  \overline{\mid C (\tau_1)- C (\tau_2) \mid }, 
\end{equation}
where overbar denotes the average over all matrix elements. We then use multidimensional scaling (MDS)~\cite{mds_2009} on these distances (see supplementary material Fig. S3) to obtain coordinates in a lower dimensional space keeping between-object distances (similarities between correlation matrices at different time)  preserved as much as possible within a given tolerance. In our case we project the distances into a three dimensional (3D) space, which allows visualization of projections or animated 3D arrangements.  We then use $k$-means clustering~\cite{teknomo2006k,jain2010data} to optimize the intra cluster distance, averaged Euclidean distance between all points within clusters, obtained using different random initial conditions. 

The optimization is carried out  varying simultaneously the noise suppression parameter for the power mapping technique as applied to the correlation matrices~\cite{Pharasi_2018,Pharasi_2019,chakraborti2020emerging} and the total cluster number. We increase the set of correlation matrices as compared to Ref.~\cite{Pharasi_2018} by choosing again $M=20$ days epoch but shifts of $\Delta=1$ day. The epoch length is slightly above the value, where noise is very destructive. The shift of the epoch in the time horizon is the shortest we can have without feeling the effects of intraday trading and with easily accessible data~\cite{Yahoo_finance}. The reduction of the shift will improve the usefulness for market participants and policymakers as there will be a daily update and better statistics. We work with a larger ensemble, i.e, we have ten times more market states than the previous study~\cite{Pharasi_2018,Pharasi_2019,pharasi_2020}. In the supplementary material, we have shown results for same epoch of length $M=20$ days but shifts of $\Delta=10$ days over the same period 2006-2019. It is obvious, in this high dimensional space, trajectories will never cross each other unless some symmetry or similar properties compel a crossing. In three dimensional reduced space, we also expect that some arbitrary trajectories will not cross. On the other hand, it is unavoidable in two dimensional space, and this will dilute the value of visual representation.

\begin{figure}[t!]
	\centering
	\includegraphics[width=0.7\linewidth]{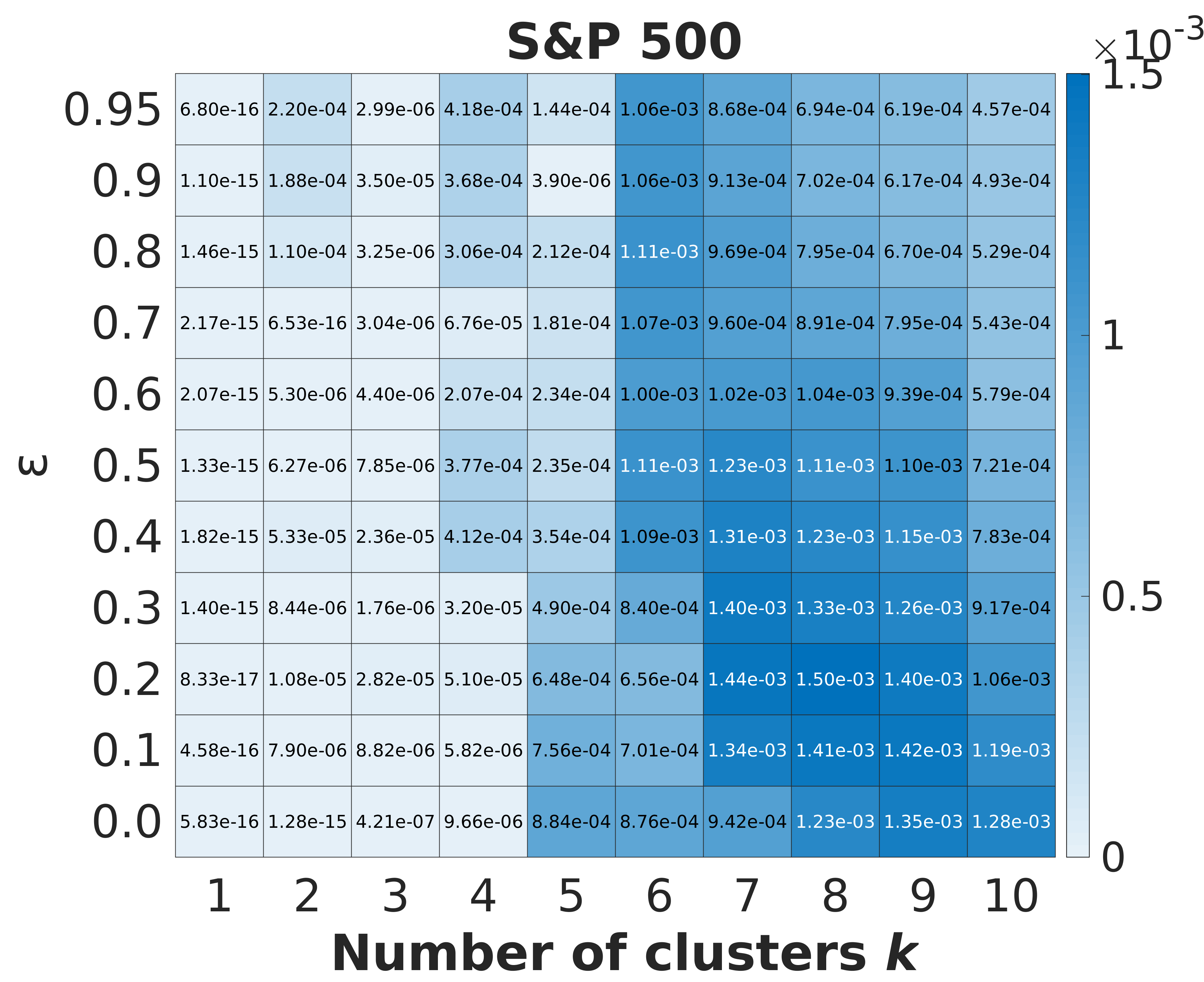}\llap{\parbox[b]{3.6in}{\textbf{{\Large (a)}}\\\rule{0ex}{2.6in}}}
	\includegraphics[width=0.49\linewidth]{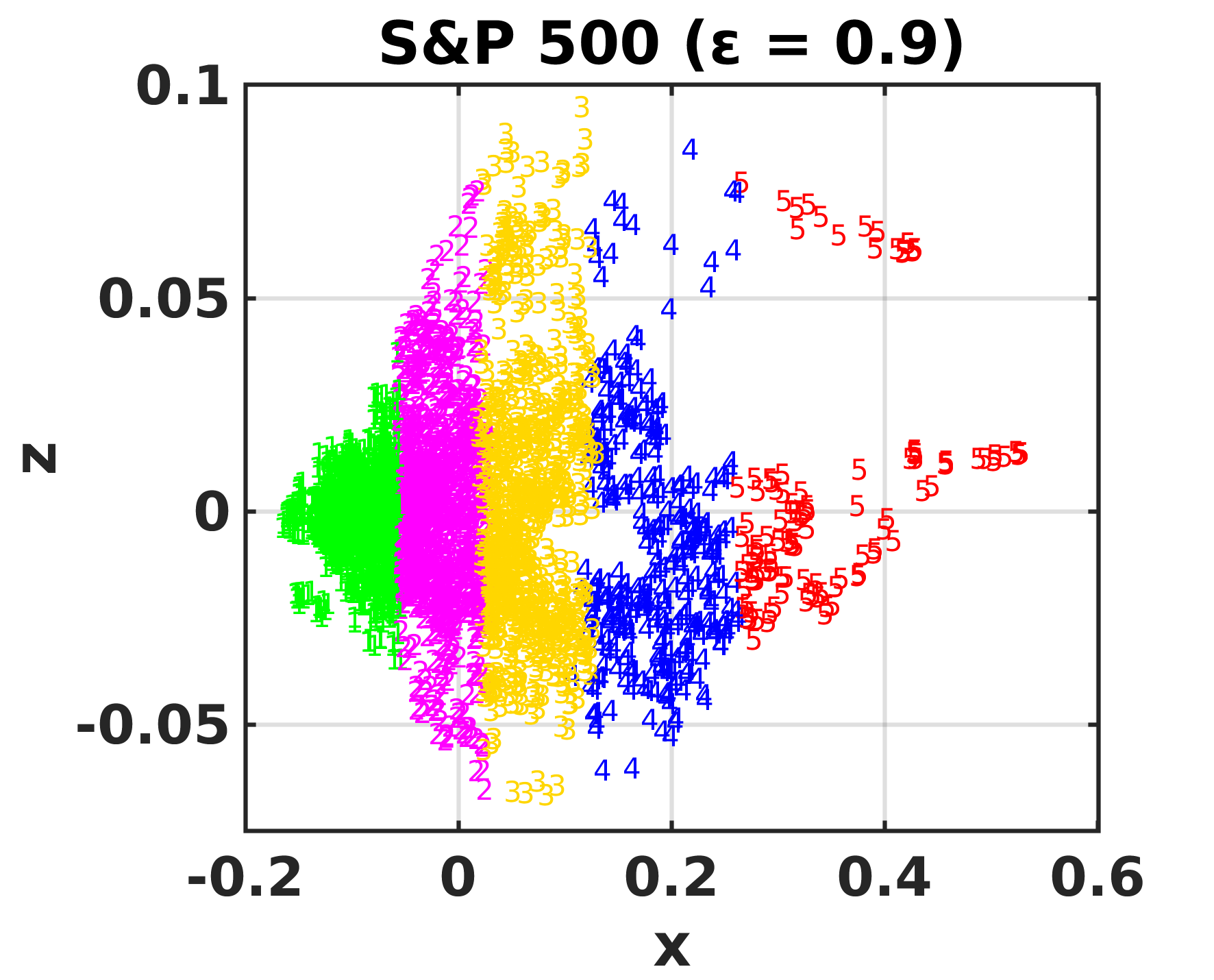}\llap{\parbox[b]{2.4in}{\textbf{{\Large (b)}}\\\rule{0ex}{1.8in}}}	
	\includegraphics[width=0.49\linewidth]{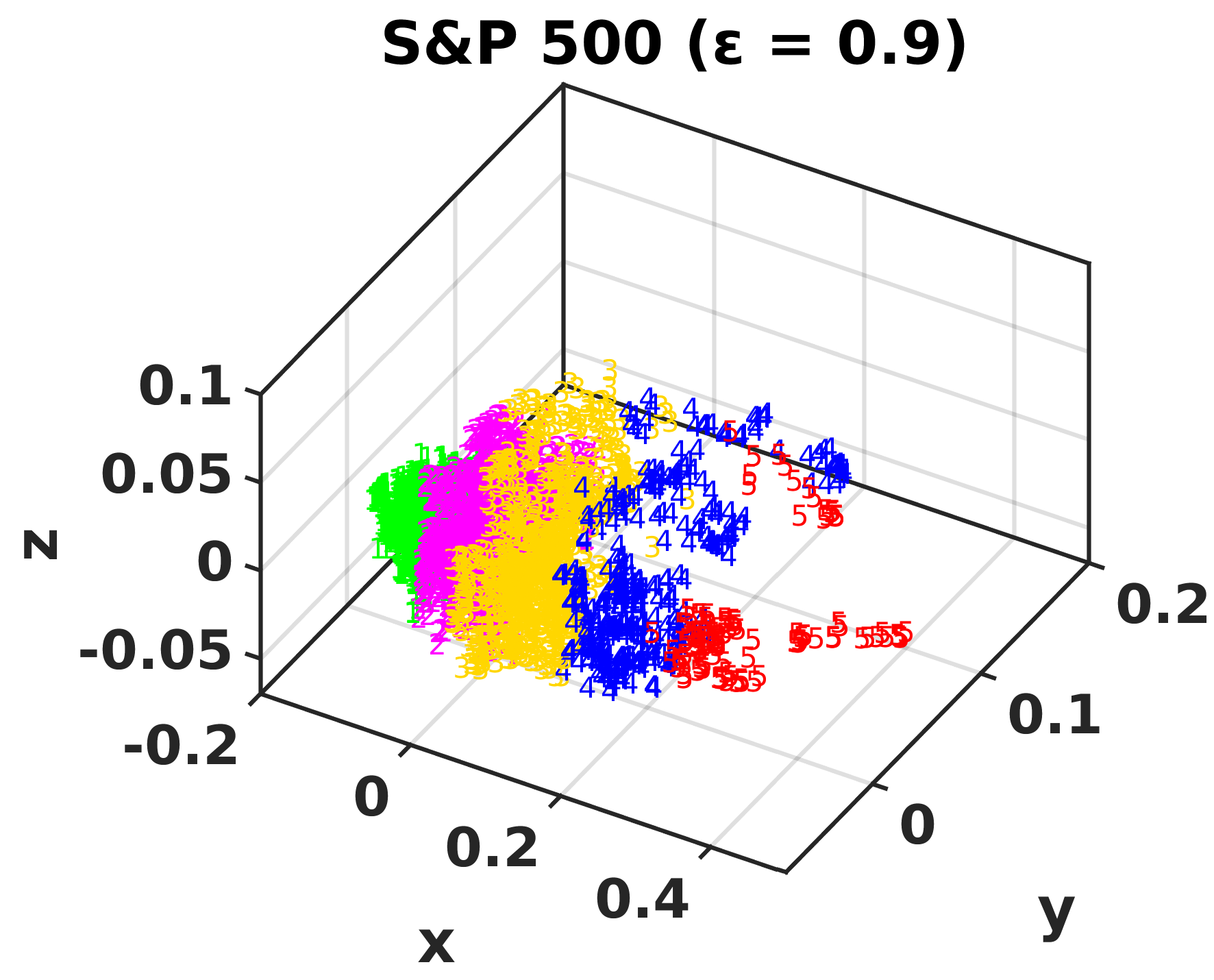}\llap{\parbox[b]{2.4in}{\textbf{{\Large (c)}}\\\rule{0ex}{1.8in}}}
	\caption{Classification of market states based on minimum standard deviation of intra cluster distances $\sigma_{d_{intra}}$ for S\&P 500 market. Plot (a) shows the measure of $\sigma_{d_{intra}}$ (colorbar) for different number of clusters $k$ and noise suppression parameter $\epsilon$. Here we use 1000 different initial conditions for the calculation of $\sigma_{d_{intra}}$. For $k\geq 4$, the minima of $\sigma_{d_{intra}}$ appears at $k=5$ and $\epsilon=0.9$. Plot (b) shows the 2D projection of (c) 3D $k$-means clustering on $xz$ plane performed on distance matrix $\zeta$ computed for $3503$ noise-suppressed ($\epsilon=0.9$) correlation matrices of S\&P 500 market with $k=5$ clusters over the period of $2006-2019$. Here, we show best 2D projection of 3D plot.}\label{intracluster_usa}
\end{figure}
\begin{figure}[ht!]
	\centering
	\includegraphics[width=0.7\linewidth]{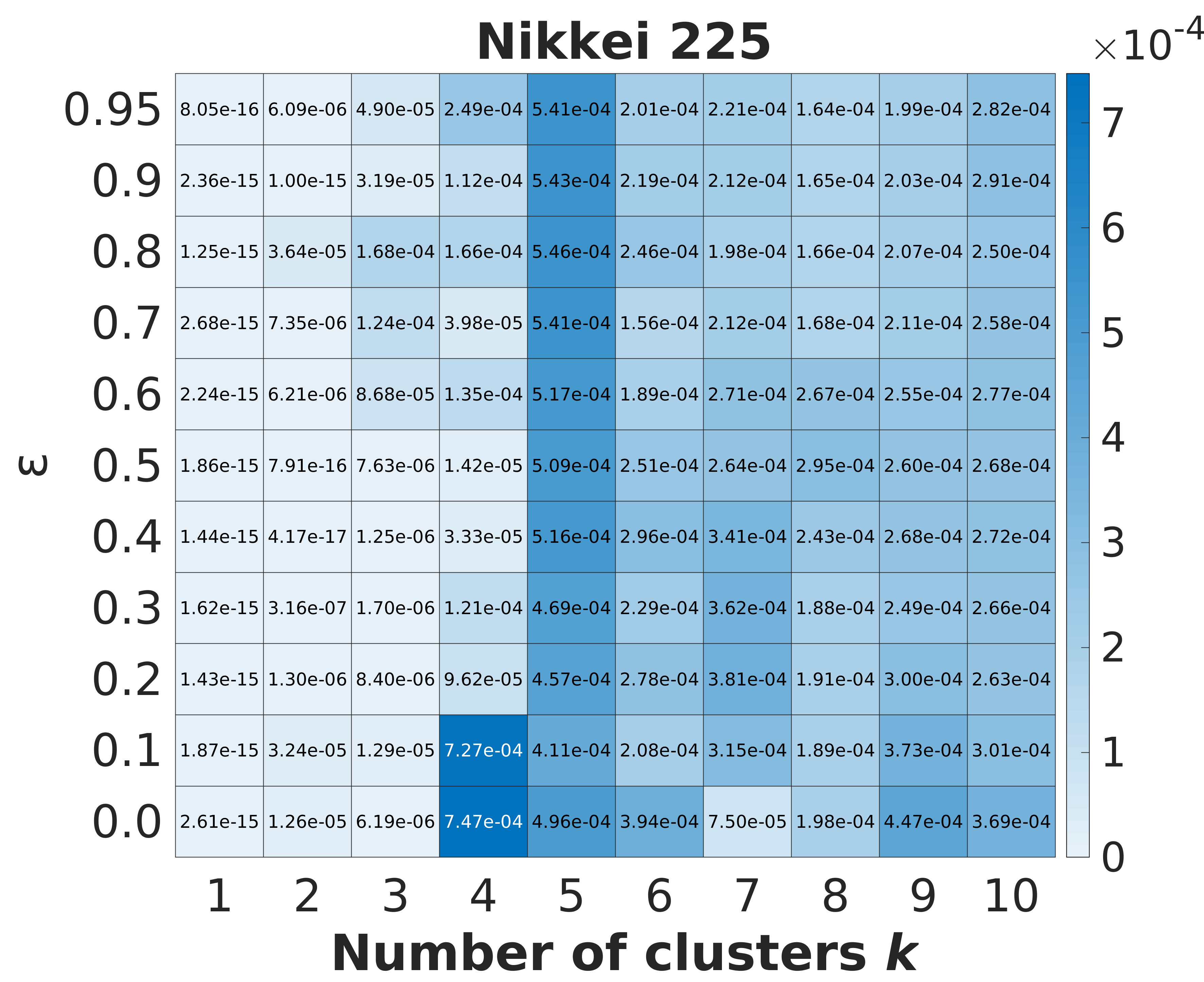}\llap{\parbox[b]{4.in}{\textbf{{\Large (a)}}\\\rule{0ex}{2.4in}}}\\
	\includegraphics[width=0.49\linewidth]{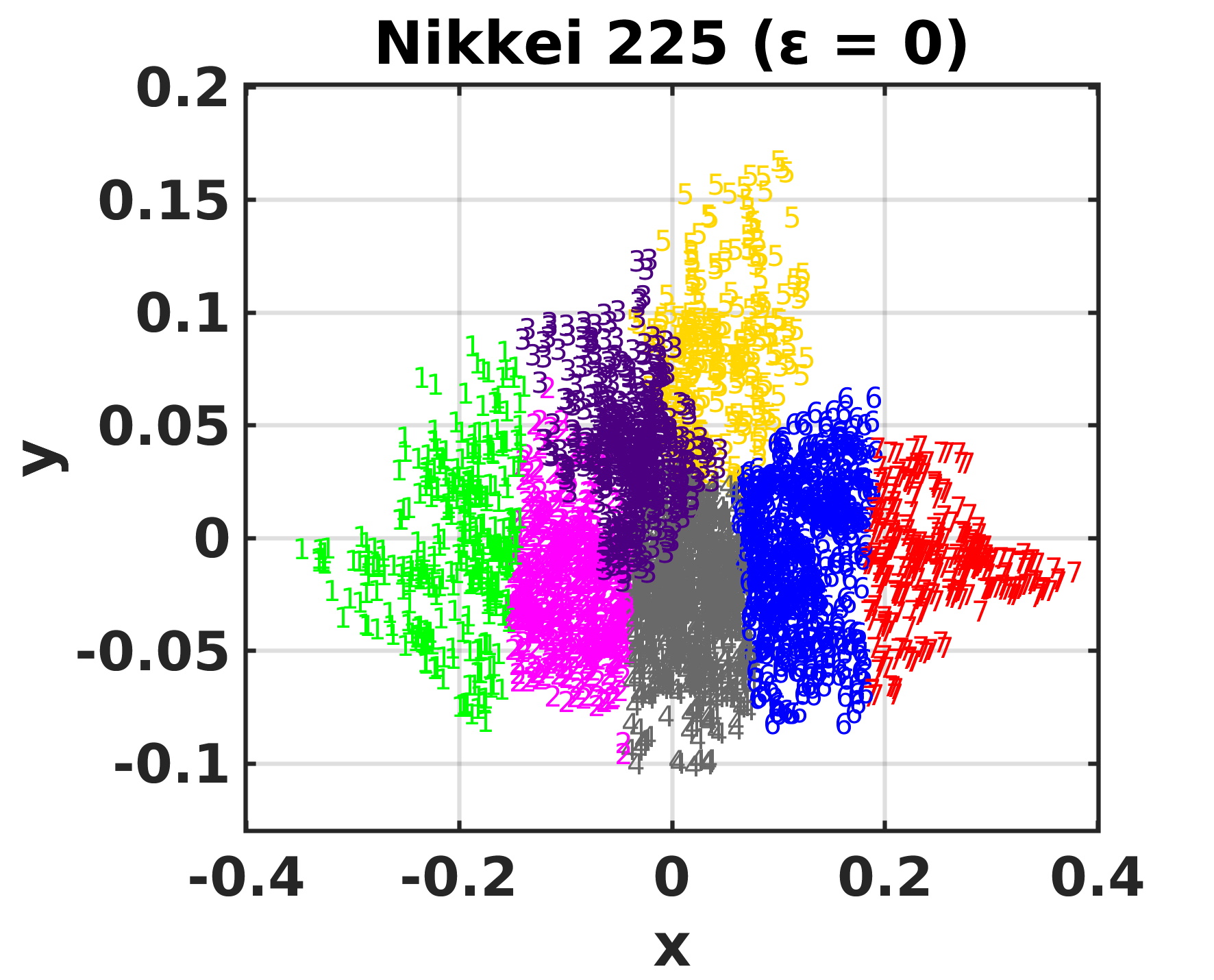}\llap{\parbox[b]{2.4in}{\textbf{{\Large (b)}}\\\rule{0ex}{1.8in}}}	
	\includegraphics[width=0.49\linewidth]{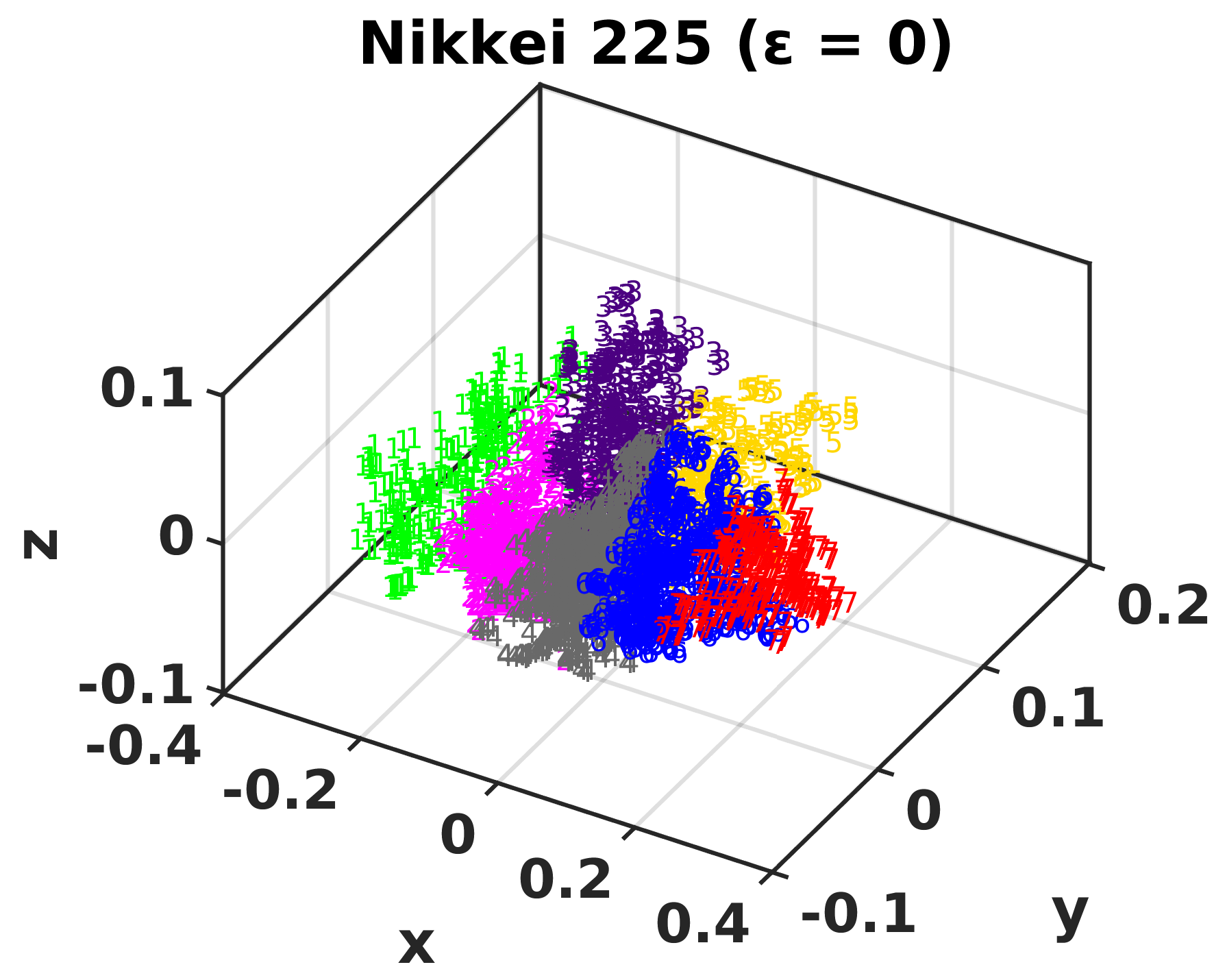}\llap{\parbox[b]{2.4in}{\textbf{{\Large (c)}}\\\rule{0ex}{1.8in}}}
	\caption{Classification of market states based on minimum standard deviation of intra cluster distances $\sigma_{d_{intra}}$ for Nikkei 225 market. Plot (a) shows the measure of $\sigma_{d_{intra}}$ (colorbar) for different number of clusters $k$ and noise suppression parameter $\epsilon$. Here we use 1000 different initial conditions for the calculation of $\sigma_{d_{intra}}$. The plot  shows the minima of $\sigma_{d_{intra}}$, for $k\geq 4$, at $k=7$ and $\epsilon=0$ for Nikkei 225 market. Plot (b) shows the 2D projection on $xy$ plane of (c) 3D $k$-means clustering with $k=7$ clusters performed on distance matrix $\zeta$ for $3439$ noise suppressed ($\epsilon=0$) correlation matrices of Nikkei 225 market  over the period of $2006-2019$. Here, we show best 2D projection of 3D plot.}\label{intracluster_jpn}
\end{figure}
\begin{figure*}[!t]
	\centering
		\includegraphics[width=0.99\textwidth]{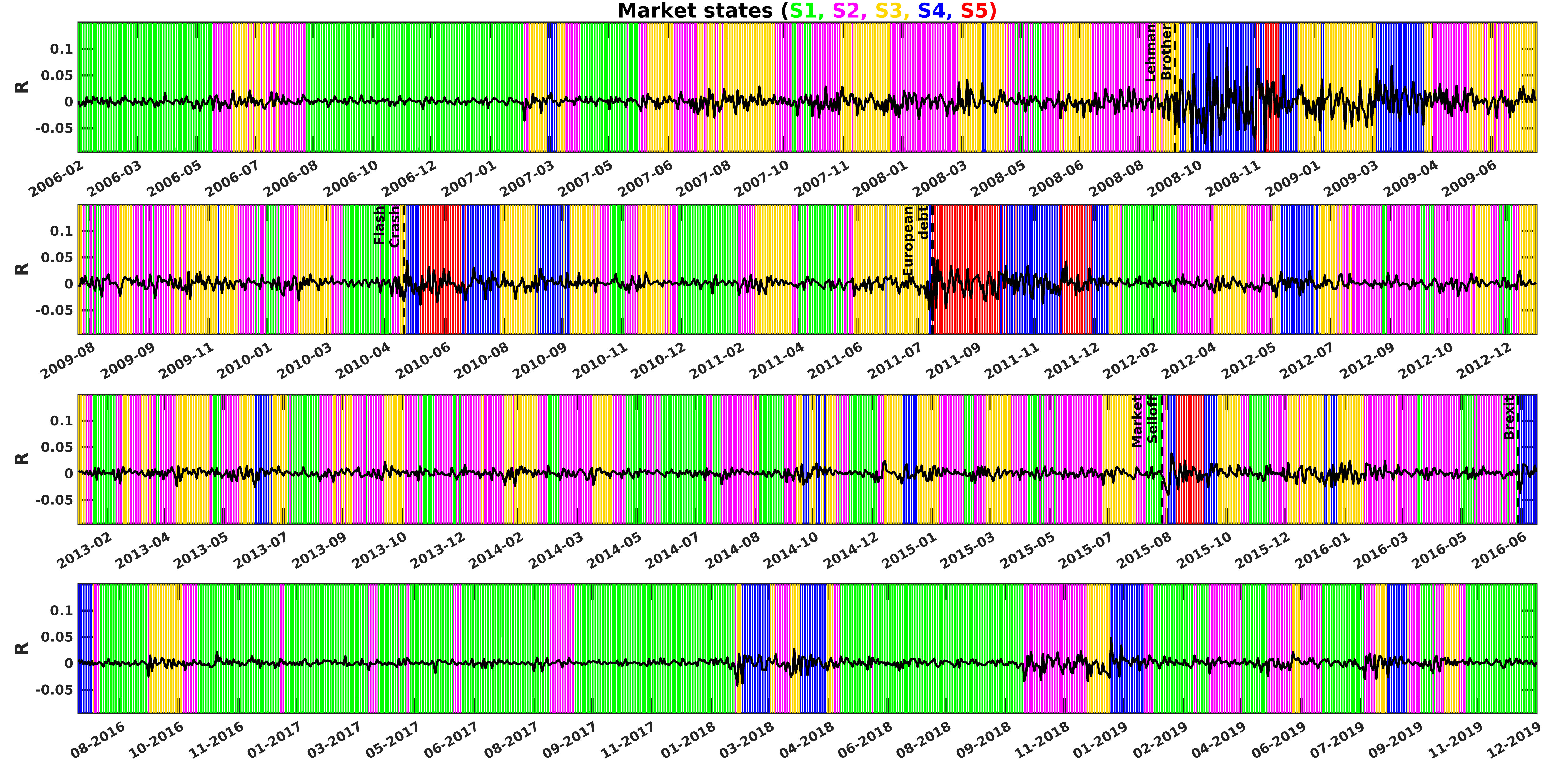}\llap{\parbox[b]{5.1in}{\textbf{{\Large (a)}}\\\rule{0ex}{2.3in}}}
		\includegraphics[width=0.99\textwidth]{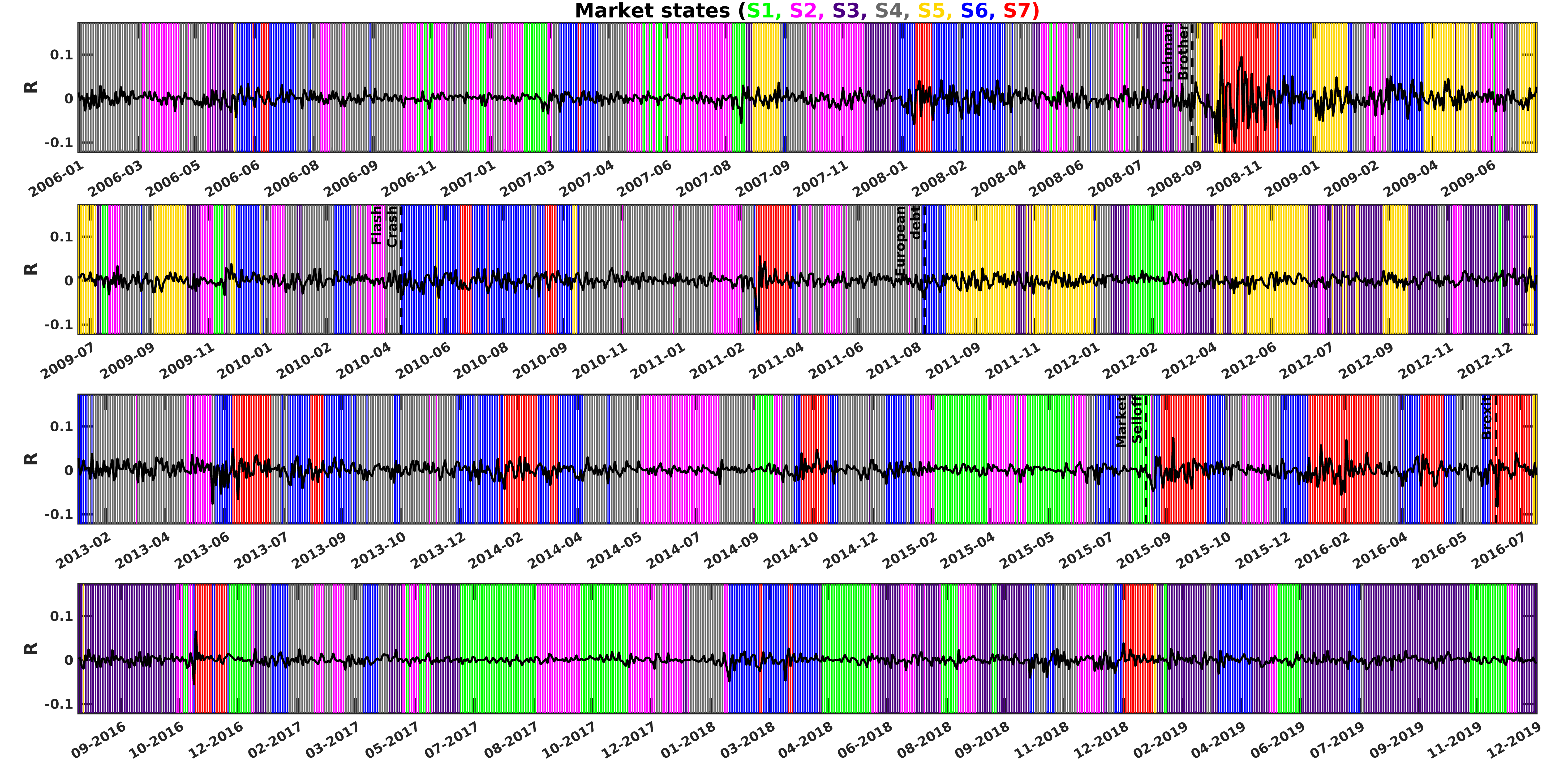}\llap{\parbox[b]{5.1in}{\textbf{{\Large (b)}}\\\rule{0ex}{2.3in}}}
		\caption{Temporal dynamics of logarithmic index return $Idx$ (black line) and the classification of the S\&P 500 and Nikkei 225 markets into their market states (vertical colored lines) over the period of $2006-2019$. (a) Evolution of S\&P 500 market through transitions among five different characterized states $S1$(green), $S2$(magenta), $S3$(gold), $S4$(blue), and $S5$(red) over the period of $2006-2019$. (b) Evolution of Nikkei 225 market through the transitions between seven different characterized states $S1$(green), $S2$(magenta), $S3$(indigo), $S4$(gold), $S5$(gray), $S6$(blue), and $S7$(red) in 14 years. Both time series are divided into four subplots (top to bottom) for clear visibility. The US market, as compared to the Japanese market, is relatively calm after 2016 and stays more in lower states. }\label{MS_evolution}
\end{figure*}
\begin{figure}[ht!]
	\centering
	\includegraphics[width=0.99\linewidth ]{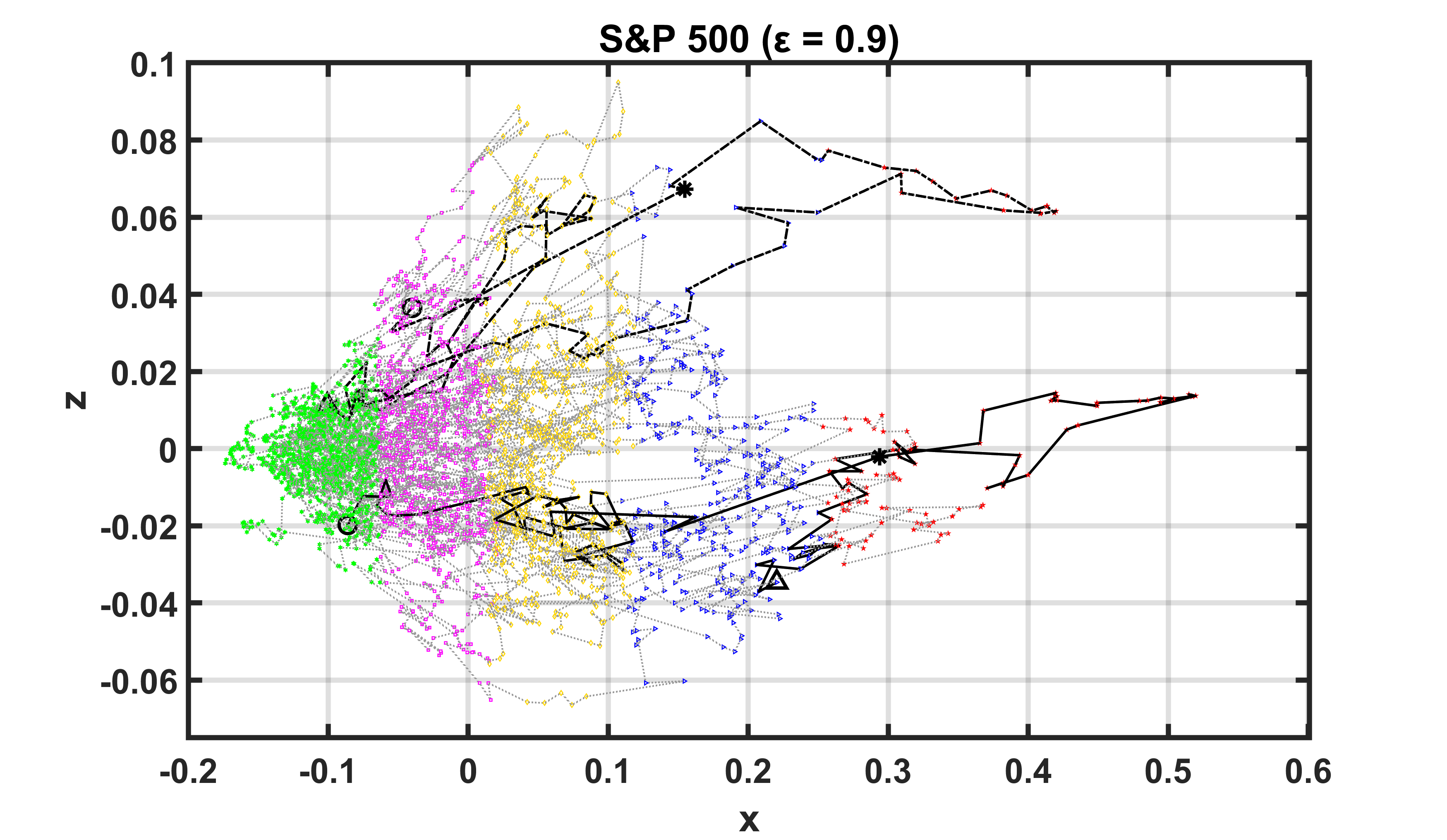} \llap{\parbox[b]{2.in}{\textbf{{\small 2015-crash}}\\\rule{0ex}{2.27in}}}\llap{\parbox[b]{2.in}{\textbf{{\small 2011-crash}}\\\rule{0ex}{1.47in}}}
		\caption{Plot shows a trajectory (grey dotted line) of the S\&P 500 market clustered into five market states $S1$(hexagram), $S2$(square), $S3$(diamond), $S4$(right-pointing triangle), and $S5$(pentagram) over the full-time horizon from  $2006$ to $2019$ constructed from epochs of $M=20$ days and shifts of $\Delta=1$ day. The trajectory, better visible in low-dense areas, is formed by connecting the points in the dimensional scaled correlation space. We find two separate branches during the evolution of the crash state: the upper branch is for the 2015 stock market sell-off crash (2015-crash) and the extreme end of the lower branch is for the August 2011 stock markets fall (2011-crash). Near these two critical events, the evolution dynamics of the two trajectories (110 days) in the correlation space are shown by a dash-dotted line and a solid line. Circle, asterisk, and triangle symbols represent the starting point, crash, and ending point of the trajectories respectively.}\label{trajectoryS5}
\end{figure}
\section{Results}\label{results}
\subsection{Identification of market states}
For the two-fold optimization, we use the standard deviation of averaged intracluster distance $\sigma_{d_{intra}}$, averaged points to centroids distance, and explore a map of noise suppression parameter versus the number of cluster $k$ taking different initial conditions in the $k$-means formalism while calculating the intracluster distance. Thus, we find a landscape that typically has various local minima. Figure~\ref{intracluster_usa} (a) shows this map for the S\&P 500 considering clusters from 1 to 10 and a noise suppression parameter $\epsilon$ from $0$ to $0.95$ with the standard deviation of intracluster measure $\sigma_{d_{intra}}$ indicated by a tone scale. 1000 different initial conditions are used for $k$-means clustering to measure $\sigma_{d_{intra}}$, and their minimum value is a measure of robust clustering. We shall not look at the results for less than four clusters ($k\geq 4$), though they are given for completeness in the picture because we considered that fewer than $4$ market states are not helpful. For the S\&P 500, the best choice seems to be $k=5$ clusters with $\epsilon=0.9$. Figure~\ref{intracluster_usa} (b) shows a projection of (c) the $k$-means clustering on $xz$-plane for five clusters. Here we show the best projection of 3D to 2D space.

Figure~\ref{intracluster_jpn} (a) shows the same color map for the Nikkei 225 of different numbers of clusters $k$ and noise suppression parameters $\epsilon$ with the standard deviation of intracluster measure $\sigma_{d_{intra}}$ indicated by a tone scale.  For the Nikkei 225, the best choice seems to be $k=7$ clusters and $\epsilon=0$. Here robust clustering is achieved without noise suppression, i.e., $\epsilon=0$. Figure~\ref{intracluster_jpn} (b) shows a projection of (c) the $k$-means clustering for seven clusters. The Japanese market shows two more clusters and they are no longer not aligned according to the increasing average correlation.

\subsection{Dynamics}
Figures~\ref{MS_evolution} (a) and (b) show the evolution for S\&P 500 through the transitions among five market states $S1$(green), $S2$(magenta), $S3$(gold), $S4$(blue), and $S5$(red) and for Nikkei 225 among seven market states $S1$(green), $S2$(magenta), $S3$(indigo), $S4$(gray), $S5$(gold), $S6$(blue), and $S7$(red) for the period of $2006-2019$, respectively.  
The type of evolution of our analysis is independent of the number of internal and external factors.

Following the recent paper~\cite{Pharasi_2018} we use $M=20$ trading days epoch with the full end-of-the-day information and shifting epochs by $\Delta=1$ day at a time. This offers a better possible view of dynamics.

We shall see in section \ref{risk} the one-day shift allows us also to follow a trajectory of the correlation matrices, best viewed after dimensional scaling to a three-dimensional space. We show the trajectory over the full-time horizon for the S\&P 500 in figure~\ref{trajectoryS5}. As we noted before, the clusters are contiguous, but the fifth cluster, in an intuitive way, would be divided into two branches: one for the upper arm and the other for the lower arm. Yet allowing for six clusters, this does not change. Rather the sixth one divides the branches further into two separate areas. And indeed, if we follow the outer part of the upper and lower arms in figure~\ref{trajectoryS5}, they correspond to one piece of trajectory each, the upper arm corresponds to the 2015 stock market selloff crash, and the extreme end of the lower arm to the 2011 European sovereign debt crisis.

\begin{center}
	\begin{figure*}[ht!]
	\centering
	\includegraphics[width=8.cm]{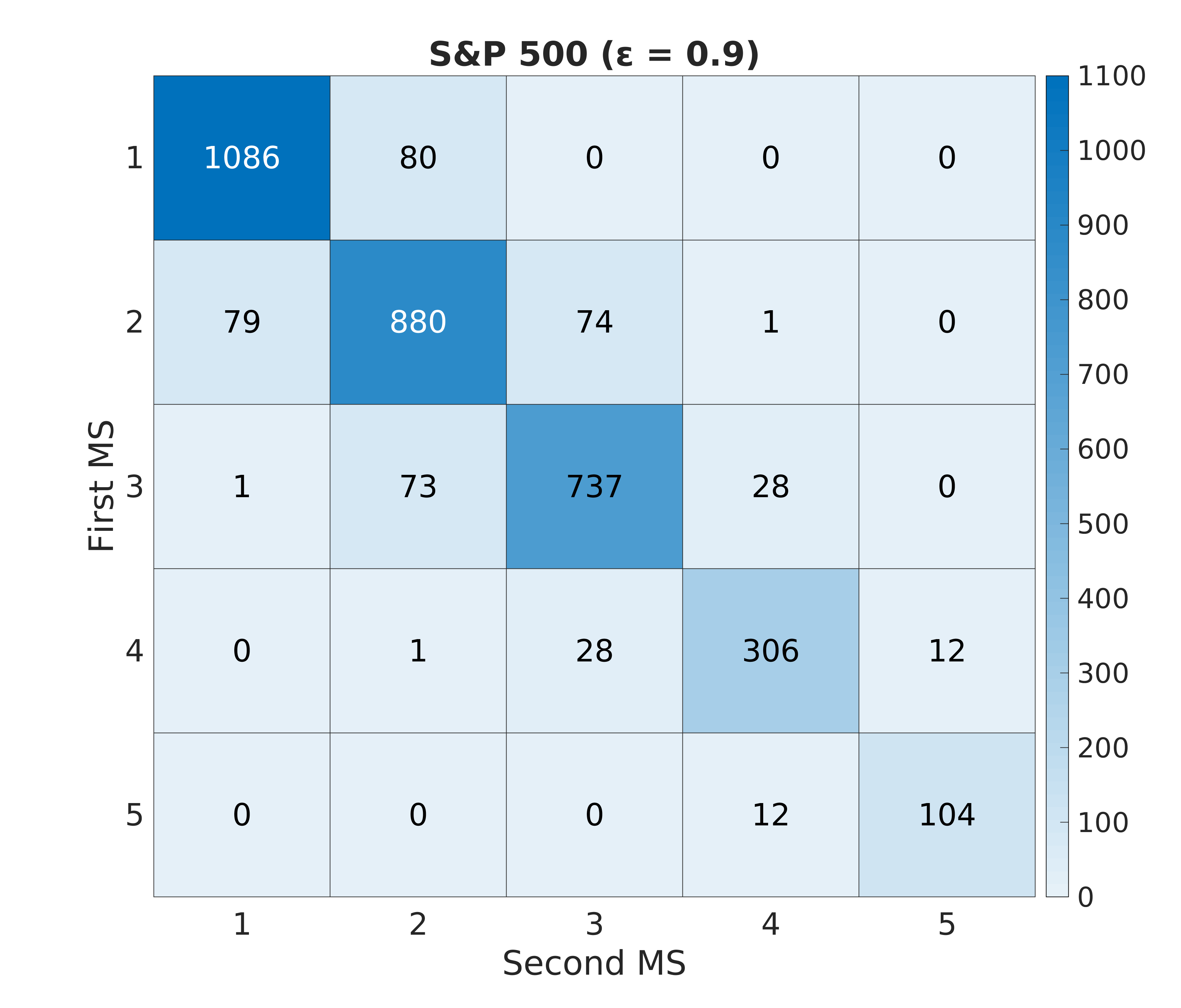}\llap{\parbox[b]{3.5in}{\textbf{{\Large (a)}}\\\rule{0ex}{2.5in}}}
	\includegraphics[width=9.5cm]{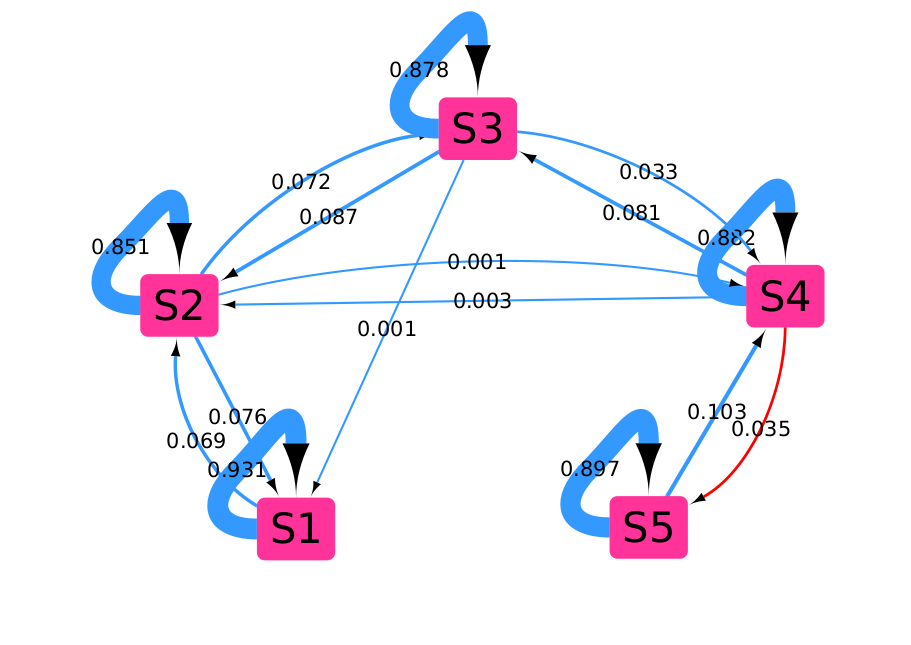}\llap{\parbox[b]{3.7in}{\textbf{{\Large (b)}}\\\rule{0ex}{2.5in}}}
	\caption{Heat map of the transition counts (frequencies) of paired market states (MS) for S\&P 500 market. The market shows back and forth transitions between the five market states. Sometimes the market remains in a particular state for a longer time and sometimes it jumps shortly to another state and bounce back or evolve further. Transitions to the nearby states are highly probable. The network plot of transition probabilities between different states of S\&P 500 is shown in (b). The probability of market state transitions from  $S4$ to $S5$ is $3.5\%$. There is no transition from lower states $S1, S2, S3$ to the highest state $S5$ which gives flexibility to hedge the penultimate state $S4$.}\label{transition_probabilities_usa}
	\end{figure*}
\end{center}
\begin{center}
	\begin{figure*}[ht!]
	\centering
	\includegraphics[width=9cm]{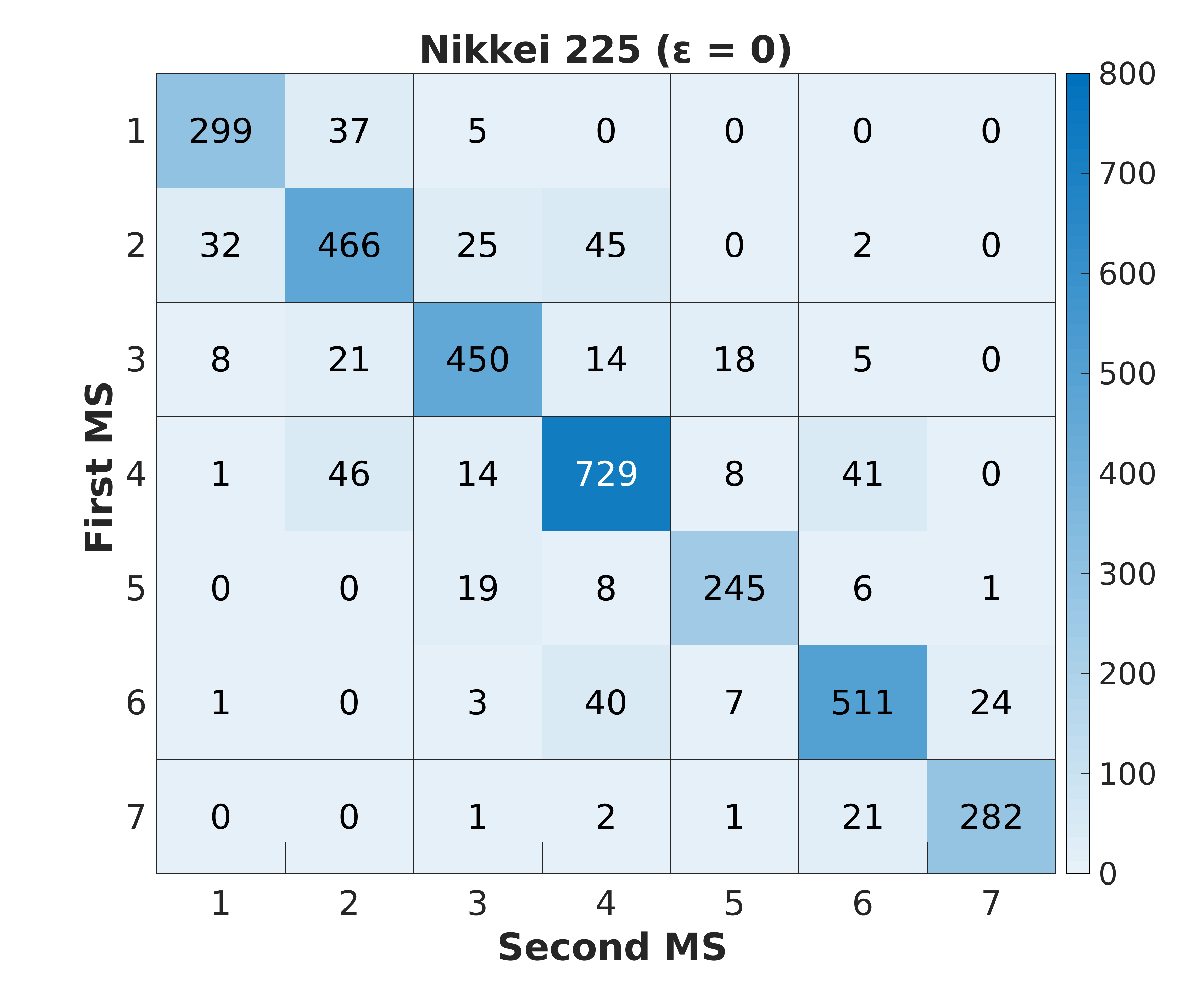}\llap{\parbox[b]{3.6in}{\textbf{{\Large (a)}}\\\rule{0ex}{2.8in}}}
	\includegraphics[width=10.5cm]{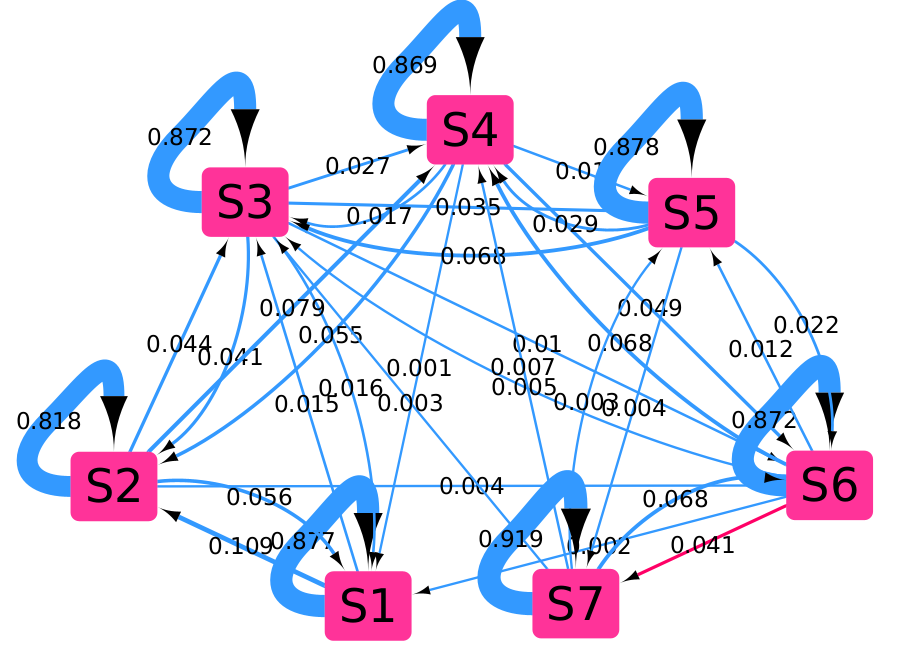}\llap{\parbox[b]{3.9in}{\textbf{{\Large (b)}}\\\rule{0ex}{2.8in}}}		
	\caption{Heat map of the transition counts (frequencies) of paired market states (MS) for Nikkei 225 market. The network plot with transition probabilities among seven market states of Nikkei 225 is shown in (b). The probability of market state transitions from  $S6$ to $S7$ is $4.1\%$. The transition counts/probabilities for $S2\rightarrow S3$ and $S4 \rightarrow S5$ are lesser than  $S2\rightarrow S4$ and  $S4 \rightarrow S6$, respectively.}\label{transition_probabilities_jpn}
	\end{figure*}
\end{center}
\subsection{Risk implications and characteristics of the method}\label{risk}
The results presented in the previous sections are for day-to-day shifts but the reader should be conscious about the fact that the length of the epoch over which correlations are calculated is $M=20$ trading days. Also, the correlation between the mean or the highest eigenvalue of the correlation matrix and logarithmic increments of the index is not perfect for several reasons: First, the index is weighted, while the correlation methods are not. Second, we only take companies that have been in the index for the entire time horizon. Third, and that is not relevant for this time interval but it is for historical data: If we have strong anti-correlations in a crash, as was the case for the petroleum industry, this depresses the average correlation somewhat; This could be compensated by using the average of the absolute values of the correlation matrix elements. Nevertheless, the movement of the index log-return $Idx$ is the central indicator of an overall crisis, while sectoral indices are likely relevant for crises limited to one or a few sectors, like the internet bubble of 2002. Typically we will enter the high correlation region rapidly if the index crashes, and thus we expect the correlation to jump upward during a crash. On the other hand, the states are roughly aligned along the axis of average correlation. For the US market, in figure~\ref{intracluster_usa} (b) and (c), we find average correlations correspond to each market state $\mu(S1, ~S2, ~S3, ~S4, ~S5) =  (0.19, ~0.32, ~0.45, ~0.57, ~0.71)$ and these states are ordered according to their mean correlation $\mu$. In the Japanese market, figure~\ref{intracluster_jpn} (b) and (c) show the average correlation corresponds to each market state $\mu(S1, ~S2, ~S3, ~S4, ~S5, ~S6,  S7) = (0.19, ~0.30, ~0.36, ~0.42, ~0.45, ~0.55, ~0.68)$ where states $S2$ \& $S3$, and $S4$ \& $S5$ are having similar average correlation. 
Market states $S2$ \& $S3$ and states $S4$ \& $S5$ could be joined pairwise reducing the cluster numbers to five states if we would base clustering on the average correlation alone. Thus the statement that clustering follows largely the average correlation~\cite{Munnix_2012} does not seem to tell the whole story for the Japanese market far away from the crash period. As we focus in this paper on the behavior near crisis, we have not yet explored how to associate it with additional parameters, maybe the Ricci curvature measure will be important \cite{samal2020network}. Ricci curvature focuses on pairwise interactions and considers higher-order relationships. This approach could unveil hidden complexities arising due to the structural changes within networks and provide deeper insights into market behavior during critical periods, e.g., bubbles, crises, etc. In general, the crash usually occurs at the boundary between the second-highest and the highest (critical) states on the way up. Externally induced crashes may or may not follow the expected pattern, because a very stable situation is less prone to get unbalanced fast if some external event tends to perturb the equilibrium. 

Thus our results suggest that a crash risk protection is necessary, basically, when entering in the second-highest state from below. This protection might have to be extended if we enter the highest state and will be suspended if we go to a lower state. The possibility of such protection would depend on the portfolio and on the price of the derivatives needed for the protection. This, in turn, indicates that we have to optimize the number of clusters if this can be done exactly or approximately maintaining zero or near to zero transitions from any of the lower states to the highest states, because it typically will reduce the number of days we remain in the state next to the critical state. We find that the robust clustering  (minima of $\sigma_{d_{intra}}$ for $k\geq 4$), as well as an approximately tridiagonal transition matrix, is obtained for $k=5$ and $\epsilon=0.9$ for S\&P 500 market. Figure~\ref{transition_probabilities_usa} (a) shows the transition counts of paired market states for S\&P 500 market. The most probable transitions between the market states are either to the same market state or to nearby market states. Figure~\ref{transition_probabilities_usa} (a) shows zero transition from $S1, ~S2, ~ S3$ to $S5$ with twelve transitions from $S4$ to crash state $S5$ and behaves as  a precursor to the market crashes~\cite{Pharasi_2018, Pharasi_2019}. Figure~\ref{transition_probabilities_usa} (b) shows the probability graph for the same. For Nikkei 225 market, the transition counts of the paired market state and probability graph are shown in figure~\ref{transition_probabilities_jpn} (a) and (b), respectively, with $k=7$ and $\epsilon=0$. Japanese market is more complex than US market~\cite{chakraborti2020emerging}. Also note that the transition probabilities of the Japanese market between states $S2$ and $S3$  are smaller than $S2$ and $S4$, and the exact counts ($S2\rightarrow S3 =25$, $S2\rightarrow S4=45$) are visible in the figure.~\ref{transition_probabilities_jpn} (a). The same behavior is noticed during the transitions between states $S4$ and $S5$, and $S4$ and $S6$. This introduces an additional criterion on which the market state classification can be optimized. We choose to make that decision after avoiding, if possible, transitions between neighbors at a larger distance and in particular from lower states to the highest one. Yet we do not intend to limit our considerations to an early warning of impending crashes. We believe that the concept of market states has a deeper sense and in the Japanese market, our wider point of view (as compared to Ref.~\cite{Pharasi_2018,Pharasi_2019}) shows a more interesting result, whose risk implications we do not know at present. A component perpendicular to the one implicit in the average correlation becomes important. At first sight, the clustering in figure~\ref{intracluster_jpn} (b) and (c) is not very obvious, but if we look at the transition matrix in figure~\ref{transition_probabilities_jpn} (a), one can clearly see that transitions between these two states are less, or in other words, there exist two visibly different passways between the lower, mainly noisy, states and the highly correlated states. It seems that, to some extent, the dynamics indicate two different routes between calm markets and the high-risk zone, and further study is required to understand this property of financial markets. 

We have also examined the stochastic nature for both of the markets using the general theory of Markov chain~\cite{ross1996stochastic} briefly described in Eqs. 1-3 of Ref.~\cite{Pharasi_2018}. The probabilities obtained for five market states of S\&P 500 are $P(S1,S2,S3,S4,S5)=(0.333,0.295,0.24,0.099,0.033)$ and for seven market states of Nikkei 225 are $P(S1,S2,S3,S4,S5,S6,S7)=(0.099,0.166,0.152,0.242,0.081,0.170,0.089)$. The actual probabilities calculated from the frequencies given in figure~\ref{transition_probabilities_usa} (a) for five market states are  $0.333, 0.295, 0.24, 0.099, 0.033$ and in figure~\ref{transition_probabilities_jpn} (a) for seven market states are $0.099, 0.166, 0.150, 0.244, 0.081, 0.170, 0.089$ for S\&P 500 and Nikkei 225, respectively. These probabilities obtained from both methods are indeed close indicating the stochastic behavior of the stock markets.
 
  Finally, we want to show, which advantages may be gained and at what price, if we artificially augment the cluster numbers (vertically) from 5 to 8 and from 7 to 8 for the S\&P 500 and the Nikkei 225 markets analysis, respectively. We see that some excitations from lower states to the highest occur, which is not useful. To compensate this disadvantage a hedge on a fewer days would be recommended. We hope this short paragraph will stress the long way from abstract theory to practice, but also may encourage portfolio managers and policymakers to look deeper into the unfolding scenario to apply it with the necessary care.
\section{Discussion and Outlook}\label{discussions}
We present improved criteria for the definition of clusters and thus market states and apply these to data sets of pre-COVID-19 pandemic era (2006-2019) for both the US and the Japanese stock market as represented by the stocks in the S\&P 500 and the Nikkei 225 indices. We have focused on the obvious need for frequent updates of data and the fact that intraday trading has its own dynamics ~\cite{admati1988theory,back1998long}, as well as we have eliminated data from the last century which seem of limited interest and influence the clustering process unduly. We thus first make a rerun of the considerations of a previous paper~\cite{Pharasi_2018} changing both the time horizon (now 2006 - 2019) and the shifts (still the length of the epoch is $M=20$ trading days) to $\Delta=1$ day. The results are now improved as the number of correlation matrices in each cluster is greatly increased as compared to $\Delta=10$ days shift~\cite{Pharasi_2018}. The number of transitions from lower states to the highest states, where crises occur, is reduced in a significant manner, which is an indication that hedging against general market meltdowns has better chances of being effective without a prohibitive cost. Yet this will depend on the price of the necessary derivatives, which will, in turn, depend on the portfolio. Also if used as an element of central bank intervention, the timeliness is improved. In the Japanese market, we find a bifurcation in the path from quiet states to the more critical ones, clearly visible in the transition matrix, and well developed in the sense that transitions between the two states, which are roughly at the same level of average correlation, are less in number. This indicates that a variable different from the average correlation acquires importance. What this variable might be and how it affects the market is an open question. Summarizing, this paper is an important step to make the methods proposed practical in the sense that they can be incorporated into a multicomponent predictive program for financial markets that might also include intra-day features, fundamental aspects, and external effects. Yet the procedure is by no means restricted to financial markets. Different aspects may get higher weights depending on the purpose; thus if taken as a policy tool the aspect of the cost of derivatives may be less important or a separate consideration altogether and correspondingly the differentiation into larger numbers of states may be less relevant. 


The work to analyze the impact of COVID-19 on the financial markets is in progress. We also hope to expand an idea given in Ref.~\cite{pharasi_2020} to model the clusters by introducing random matrix theory simply based on the observed average correlation matrix of each cluster as its representative and using it to construct a correlated Wishart orthogonal ensemble (CWOE). Another direction we mentioned above is the understanding of the stochastic dynamics of the markets and this work is in progress. 

\section*{Acknowledgment}
The authors are grateful to Anirban Chakraborti and Francois Leyvraz for their critical input and suggestions. The authors also thank CIC AC-UNAM for their hospitality at various events, DGAPA, Mexico for financial support under grant number AG101122, and CONACyT, Mexico for financial support under FRONTERAS grant number 425854.  H.K.P., P.M., and S.S. are grateful for the financial support provided by UNAM-DGAPA and CONACYT Proyecto Fronteras 952 and 425854. T.H.S. and H.K.P. acknowledge the support grant by CONACYT Proyecto Fronteras 201, UNAM-DGAPA-PAPIIT AG100819 and IN113620. T.H.S., P.M., and H.K.P. also acknowledge computing support under project LANCAD-UNAM-DGTIC-016. 
\newpage
\section*{Appendix A. Multidimensional scaling}\label{App:mds}
Multidimensional scaling (MDS) aims to map $Fr$ number of objects (correlation structure $C(\tau)$), whose pairwise dissimilarity or distance is given by $\zeta(\tau_1,\tau_2)$, into a lower dimensional space. Distance matrix $\zeta(\tau_1,\tau_2)$ would be written as:
\begin{equation*}
\zeta(\tau_1,\tau_2) =
\begin{pmatrix}
\delta_{1,1} & \delta_{1,2} & \cdots & \delta_{1,Fr} \\
\delta_{2,1} & \delta_{2,2} & \cdots & \delta_{2,Fr} \\
\vdots  & \vdots  & \ddots & \vdots  \\
\delta_{Fr,1} & \delta_{Fr,2} & \cdots & \delta_{Fr,Fr}
\end{pmatrix}
\end{equation*}
The goal of the MDS is to define the coordinates $x_1, x_2, \dots, x_{Fr}$ such that the pairwise distance $|| x_i-x_j||$ is nearly equal to $\delta_{ij}$ upto a given tolerance limit, i.e., $|| x_i,x_j|| \simeq \delta_{ij}$ for all $i,j\in 1,2,\dots,Fr$, where $|| . ||$ is the vector norm and Euclidean distance between objects. It is important to know that the choice of $x_i$ is not unique and an arbitrary transformation, rotation, and reflection cannot change the pairwise distance $|| x_i,x_j||$. The choice of 2D and 3D is good for better visualization.

\section*{Appendix B. $k$-means clustering}\label{App:kmeans}
We use $k$-means clustering or Lloyd’s algorithm to partition the $Fr$ data points, obtained from MDS map, into $k$ clusters defined by centroids. The algorithm consists of following steps
\begin{enumerate}
\item Choose randomly $k$ initial cluster centroid for $Fr$ data points.
\item Compute point-to-centroid distances of all observation points.
\item Assign each point to the cluster with the nearest centroid.
\item Calculate the new centroid by averaging the observation in each cluster after update.
\item Repeat the steps 2 to 4 until cluster assignments or the centroid positions of clusters do not change.
\end{enumerate}
We use minima of the standard deviation of intracluster distance  $\sigma_{d_{intra}}$ for optimizing the number of clustering $k$ for $k$-means clustering. The intracluster distance $d_{intra}$ is an average of averaged Euclidean distance of the point-to-cluster distance of each cluster. The standard deviation $\sigma_{d_{intra}}$ of 1000 initial random conditions is then calculated for each $k$ value. The $k$ value at which  minimum of the $\sigma_{d_{intra}}$  is attained is the best choice of the number of clusters $k$ value. The minimum of the $\sigma_{d_{intra}}$ is  corresponding to the robust and stable clustering and calculated from the similar data points in each cluster using 1000 different conditions.  Here we are mainly focused on $k\geq4$.

\newpage
\bibliography{NJP_2020}
\end{document}